**Manuscript title:**

**Detecting of Low Energy Interactions and the Effects of Energy Accumulation in Materials.**

**Manuscript corresponding author:**
**Sergey Pereverzev**

**Address:**
**Lawrence Livermore National Laboratory,**
**7000 East Avenue,**
**Livermore, Ca 94550, USA**

**E-mail: pereverzev1@llnl.gov**
**Alternative E-mail: persev@gmail.com**

**Authors:**
**Sergey Pereverzev**
**Affiliation: Lawrence Livermore National Laboratory**

**Detecting Low-Energy Interactions and the Effects of Energy Accumulation in Materials.**

**Abstract**

**Elusive as dark matter particles are, they are not the only entities that can produce small energy releases deep inside the most sensitive detectors. Cosmogenic and residual radioactivity as well as other factors can produce and slowly accumulate long-living excitations and defect configurations in materials. Unsteady and avalanche-like releases of accumulated energy can limit the dynamic range of detectors and the sensitivity of experiments. This type of mechanism, while not widely discussed in the dark matter and neutrino detectors community, got a lot of attention in discussions about systems with energy flow in the context of non-equilibrium thermodynamic. This paper explores avalanche hypothesis using published data on low-energy background in detectors and available condensed-matter physics information on excitations, defects, and energetic molecules which can be produced in detector materials and examine ways to use these models to understand the origin of low-energy background in dark matter and other detectors and for suppression of this parasitic background.**

**Introduction:**
Significant interest in the dark-matter problem and physics beyond the Standard Model has encouraged experimenters to "turn all stones" [1, 2, 3], including searching for light-mass dark matter particles at the limits of low-energy detector's sensitivity [4, 5, 6, 7, 8]. While these searches show promise, they suffer low energy background effects that change with the cumulative ionization load, and which can vary with impurities and/or changes in pressure and temperature. Furthermore, troublesome correlations—such as time- and position-correlations with real events and delayed simultaneous observations of several electrons or photons after large events—appear in signals from a variety of detector types and materials. These correlations suggest that there may be unexpected, condensed matter or chemical processes and effects affecting data in the search for the elastic scattering of low-energy neutrinos on nuclei and in the quest for dark matter particles and physics beyond the standard model.

For dual-phase noble liquid detectors, these small background signals consist of approximately 1-8 detected ionization electrons, which lie in the signal range expected for coherent elastic scattering of neutrinos on nuclei (CEvNS) for low-energy reactor antineutrinos scattering on Xe or Ar nuclei [9, 10]. CEvNS has the largest cross-section for interactions known for low-energy neutrinos, so it is interesting to use this process to obtain direct experimental information on neutrinos in the range of fundamental and practical problems. Such efforts include monitoring the solar neutrino flux, detecting neutrinos from a supernova, and harnessing applied neutrino physics, including applying neutrino detectors to reactor monitoring. This yet unidentified low-energy background in *underground* experiments with Xe (e.g., Xenon1Ton [4]) and Ar (DarkSide-50 [6]) is sufficiently low to allow detection of CEvNS at a power reactor with a short (~25 m) baseline detector- would both detector and reactor will be placed deep underground. However, for detectors operating at the surface, where the cosmogenic ionization load is larger, this low-energy background exceeds expected CEvNS rates under realistic conditions (distance from the reactor and reactor power) by several orders of magnitude. Thus, the effects discussed here are not small, and resolving the observable excesses in low-energy backgrounds would facilitate progress in multiple experiments, where these effects blur data and impede discoveries.

This paper discusses mechanisms known to exists in detector materials but that have not yet been considered in analyses of low-energy background data. We propose a model that includes the production of energy-bearing states (excitations), their accumulation, and their avalanche-like annihilations, and we predict the avalanches will yield bursts of photons, electrons, quasiparticles, intermittency, and other "noisy," irregular dynamic effects. The model follows the scenario of self-organized criticality (SOC) [11,12] and predicts that this noise will increase with increasing ionization load or with energy dissipated in detector materials by any other process. We anticipate that when these underlying processes in the system are properly understood, the model may be exploited to suppress or discharge these parasitic backgrounds.

.

**Recurring patterns in the near-threshold background (*background on noise*)**
In the course of reviewing the low-energy response and background observed within a range of different particle detectors, we found a striking commonality: the number of background events rises sharply for energies approaching those required to produce photon or electron in the detector materials [6,8,13,14,15,16,17,18]—see Figs. 1, 2. In dual-phase (liquid-gas) Ar and Xe detectors, the events' histograms manifested a large number of single electrons in the detector as well as multiple-electrons events—with two, three, four, …up to eight electrons observed (as in Fig 1a). The events of these one-to-few electrons were detected during the normal low-energy event time window and looked like typical low-energy ionization events (see Fig. 1 a, b, c). Scintillator [16,18] and semiconductor [8] detectors manifested similar growth within their low-energy background, though the very low-energy portions of these detectors' events spectra—i.e., near single-electron or single-photon events—are below their current amplifier noise or PMTs background, so only events with multiple electrons/photons could be analyzed (see Fig. 1d, Fig. 2b). Additionally, when readouts in CCD devices allow single-electron sensitivity and when frames containing large ionization events were excluded, one-, two-, and three-electron background events were observable in the cold silicon CCDs [17]. Such patterns suggest a possible underlying cause justifying further examination.

The dual-phase detectors are, as a rule, designed specifically to reduce such low-energy noise and have had impressive success in this regard. Underground assembly and deployment are often employed to reduce direct cosmogenic backgrounds and induced radioactivity, and high levels of purification are essential to the performance of many of these detectors. Self-shielding and so-called fiducialization—deliberate exclusion of events identified in the outermost shell of the detection medium—are employed to inhibit the ingress of low-energy gamma or beta particles deep inside detectors. Low-angle Compton scattering of high-energy gammas and low-energy beta-active impurities in the materials should yield a practically flat or even slightly decreasing (toward zero) low-energy spectra of events inside the detector (as modeling of backgrounds in [6] shows). Thus, the operation of these detectors containing highly purified target media in low-background environments greatly challenges any notions of coincidental multi-electron events—especially when the observed rate of single-electron events is too low to explain multiple coincidences of single-electron events.

In dual-phase detectors, the parasitic electron emission is stronger after large ionization events (decay of this emission with time was the subject of multiple studies, including recent work from Purdue [19]), so data acquisition is usually vetoed for some time after muons or other large events in dark-matter particle searches; still, multiple-electron emission events leak into the final spectra, as depicted in Fig. 1b, c. In above-ground operations, the low-energy background spectrum for different types of detectors has a similar shape but is much stronger than for underground operations (Fig. 2, 3). Consistent with the dependence on depth (and thus muon exposure), low energy background increases with the total ionization load. Dependence on ionization load was also observed (Fig. 3) with the small dual-phase Xe detector available at Lawrence Livermore National Laboratory (LLNL). A recent paper from Purdue University [19] devoting to studies of parasitic background in small dual-phase Xe detector operating above ground also points out an excess of multiple electron emission events over the random coincidence of single-electron events. Some universal material mechanism that depends on the total ionization rate in the detector appears to be present that is responsible for at least part of the low-energy background in these detectors.

While it is expected that ionization energy or single UV photon energy—not to mention few-photons or -electrons events—are too large to be caused by thermal fluctuations at the detector operation temperature, this expectation assumes that detector materials are in thermal equilibrium, which may not be the case if "hot," long-living excitations with lifetimes longer than the time in between ionization events can be produced in materials.

**Self-organized criticality as a model for few-carriers background.**
The observations about background patterns resemble a specific dynamic process known as self-organized criticality. SOC initially appeared in computer simulations of large ensembles of interacting particles [11,12]. It is well-visualized in a sand pile settling through a series of small avalanches after enough sand has been slowly added to the top of the pile. Similarly, when energy is slowly pumped into and accumulates in a material with complex internal interactions, relaxation can occur in the form of avalanches. Here, the term SOC describes a process wherein interacting excitations slowly accumulate inside a system and annihilate in avalanche-like events. Several examples of SOC are known in condensed-matter physics, such as crack formation in materials under increasing stress [20] and dynamics of quantized vortexes as magnetic flux penetrates type II superconductors [21,22]. We posit those similar processes are at least partially responsible for background events in particle detectors and, possibly, quantum sensors operating at low temperatures.

What makes this notion plausible is that when avalanches are present, in many cases they lead to *specific* correlations and effects in the systems [11, 12]. For histograms of SOC-induced events, the probability of large-energy relaxation events decreases with energy not exponentially but rather polynomially (which means that catastrophic relaxation events are possible). The frequency dependence of the noise power spectral density (more applicable to superconducting sensors such as SQUIDs or bolometers) is close to 1/f.

Consistent with our hypothesis is the notion that large-energy relaxation events can be suppressed by forcing relaxations on small scales—akin to placing a sandpile on a vibrating plate to induce more small avalanches instead of allowing a few large avalanches. This technique is used to deal with snow avalanches on skies resorts- and can potentially be used with detectors.

**Energy accumulation in solids subjected to radiation and possibilities for avalanche relaxations.**

It has long been known and widely applied—see [23] for example—that exposure to ionizing radiation causes materials to store and accumulate energy, with the resulting thermally stimulated luminescence (TSL), thermally stimulated electron emission (TSEE), and thermally stimulated conductivity (TSC) appearing in bulk dielectrics, semiconductors, and organic substances as well as in films on metal surfaces, though mechanisms of energy transfer to photons and electrons emitted from the sample surface in some systems remain unclear [24]. Luminescence, electron emission, and conductivity can also be triggered by exposing irradiated materials to visible or IR light or by imposing mechanical stress [23]. Increasing the temperature of irradiated materials in a series of small steps will liberate ions/carriers from different traps, prompting their recombination with ions of opposite signs, which may lead to luminescence. Counting photons emitted during each temperature step will provide a spectrum of the ion/electron traps in the material [23]. While this description is simplified, it is important to mention that in practically all materials exposed to cosmogenic and residual radioactivity, energy will slowly accumulate in the form of trapped ions and other long-living excitations. There is a limit on the amount of energy that can be accumulated in the material. For example, dose-response curves of TSL-based dosimeters are non-linear and demonstrate saturation [23]. As the temperature increase required to observe TSL or TSEE in many common materials are relatively low (of the order of 100 C), energy barriers to trigger relaxation of individual sites are also low, and avalanche /chain-like relaxation processes may be possible.

Photon bursts have been observed in NaI(Tl) after exposure to UV light. This behavior is critical for the application of scintillator crystals for particle detection. A Saint-Gobain technical note [25] on the scintillation crystals observed that "with mild UV exposure, several pulses per second can be seen in the 6-10 keV region of a spectrum. If the crystal is stored in a dark area, this mild UV exposure will eventually disappear, although it may take from several hours to several days for the effects to stop."

This observation is consistent with our main hypothesis: *Due to the small activation energy needed to release stored energy from trapped charges, defects, or other metastable excitations, and because of the effects of clustering of defects, impurities, and trapped charges, in all materials where TSL, TSEE, or TSC are present, avalanche-like relaxation events that create bursts of electrons or photons, or bursts of conductivity can also be present.*

In alkali metal halides—such as NaI(Tl) scintillators—the presence of TSL is documented [26] and often cited as the source of non-linearities or delayed luminescence. In solid Ar and Xe, both TSL and TSEE are present [27,28]. In common semiconductor detector materials like Si and Ge not only TSC is present, but , but exoelectron emission following irradiation or mechanical processing is known [29].

Energy storage in solid materials can extend to days (as has been seen in solid Xe or Ar) and years, and there is some evidence that oppositely charged ions can stay trapped close to each other [23]. Stable positive and negative ions (molecular ions, often called charge-transfer excitons) may be produced and trapped in alkali-halides crystals (in color centers, F centers, H centers, etc.) [30, 31]. In rare gas solids (RGS), only positive native molecular ions are present—$Xe_2^+$, $Ar_2^+$, $Ne_2^+$, $He_2^+$. Impurities are therefore a likely requisite for long-time storage of negative ions in solid Xe and Ar. To explore such cases, initial experiments on energy storage in solid Xe used halogen-doped samples [32, 33, 34]. Later studies of TSL and TSEE effects in nominally pure solid Xe and Ar revealed correlations with the presence of oxygen [27,28]. Apkarian [34] observed that irradiating halogen-doped solid Xe with UV light produces short-lived so-called 'exciplex' complexes of $Xe_2^+ X^-$ (where $X$ represents Br, Cl, or F). These complexes subsequently fluoresce at sub-microsecond scales and will occasionally (about one in $10^5$ events) separate into closely spaced pairs of $Xe_2^+$ and $X^-$ ions—a process called photolysis—with the resulting ions pairs appearing as metastable excitons experiencing a wide distribution of lifetimes. Other studies revealed the presence of reaction intermediates in solid rare gases under irradiation; these neutral and charged molecules consisting of rare gas atoms, halogen atoms, oxygen, and hydrogen have a short lifetime in the gas phase but can be stabilized by a solid matrix [35].

Defect and impurity clustering occur for many reasons. Ions and defects resulting from ionizing radiation will yield inhomogeneous initial distributions of excitations in the material (i.e., along particle tracks). Alternatively, the diffusion of defects and impurities out of single-crystal grains and their accumulation on grain boundaries and around other defects—e.g., dislocations, dislocation loops, boundaries, and interfaces—will trap ions and defects within the media.

Importantly, a small temperature increase will yield TSL in most common detector materials, i.e., the energy threshold for ion de-trapping/ion-pair recombination is low—below a fraction of an eV. Once initiated, the lattice relaxations after the recombination of an ion pair can produce local lattice deformations and phonons, which, in turn, can liberate another ion. As the clustering of defects and impurities are common effects, such relaxations can prompt avalanche-like recombination, especially of ions and other energy-bearing states stored within clusters. These behaviors match our hypothesized SOC-like dynamics in these materials.

The original SOC model [11,12] has universal features resembling phase transitions, such as the absence of a characteristic time and size scale for the region where an avalanche can take place, the specific shape of the energy spectrum of events, and other features. Cluster properties can be dependent on cluster size, composition, and specific chemical interactions. We suggest the term "SOC-like dynamic" to indicate the variability in behavior for different materials.

It is important to mention here that TSL and TSEE effects in solids are strongly dependent on the number of lattice defects in materials, and both formation and recombination of molecular ions inside a crystal lattice may produce lattice defects—even when the initial ionization was caused by UV light. Thus, relaxational processes will be dependent on the number of ions/excitations present and the history of the sample (as illustrated in Fig. 4 b); this suggests we are dealing with glass-like (disordered solids) behavior.

The dynamics occurring alongside metastable excitations in particle detectors are more complicated than in the case of energy steadily pumped into a system. Energetic particles can both produce and destroy the states responsible for energy storage. The low-energy nuclear recoils (below the ionization threshold)—such as those caused by the scattering of low-energy solar neutrinos—produce hot phonons and can trigger releases of stored energy but cannot produce new energetic excitations (new ions). Furthermore, tunneling, thermal excitations, and environmental factors—such as thermo-mechanical stress or ambient pressure variations—can also trigger relaxation events.

Thermal cycling and mechanical stress/ deformation also can lead to energy accumulation and storage in materials; we will discuss later that relaxational dislocation motion can cause photon and exaelectron emission. At low temperatures, energy can be stored in much smaller portions- like energies of interactions of electron and ion magnetic moments or energies stored in quantized vortexes in superconductors and superfluids, textures of the order parameter, etc. While these energies are low, avalanche relaxation events can result in energy releases significant concerning temperature and cause unwanted and somewhat unexpected effects in detectors and quantum information devices.

We acknowledge that the SOC-like dynamic is not a universal rule in systems with energy flow and may not be the leading source of noise and background. Still, when the discovery of dark matter particles or other new physics is at stake, dynamic effects in a driven system (i.e., non-equilibrium thermodynamics effects) must be considered.

**Phenomenology to look for.**

Even with very limited knowledge of the involved microscopic processes, some interesting qualitative predictions can be done using the assumption of SOC-type dynamics in scintillators like NaI(Tl). Any ionizing radiation can cause effects similar to those produced by irradiation with UV light: after irradiation, one should expect to see an increase in the number of low-energy events looking like irradiation with 1-10 keV electrons; these signals should go away after some time after radiation exposure. It is known that exposure to red or IR radiation is leading to the suppression of TSL in alkali halides, including NaI(Tl) [31]. Likely, red and IR light are causing quenching of metastable energy-bearing states in these materials, so subsequent heating will not cause energy release and UV light production. If the same energy-carrying states are involved in avalanche-like energy relaxation/release events, exposure of NaI(Tl) to red or IR radiation after exposure to UV light should alleviate the effect of exposure to UV light, i.e., make the number of 1-10 keV – like events smaller or even completely suppress increase of background in this part of the event spectrum. If the wavelength of quenching IR radiation can be chosen outside of sensitivity limits of PMT or if an efficient blocking filter for this IR /red light can be used, continuous IR irradiation can suppress (at last partly) parasitic background caused by SOC-like dynamics in 1-10 $keV_{ee}$ energy range. Excitation quenching may be non-radiative or radiative (without or with photons emission by the scintillator). Quenching can be also caused by a mechanical strain. So, the effect of UV light likely can be alleviated by the application of ultrasound or low-frequency oscillations of hydrostatic pressure (all-directional compressions of material).

David Nygren estimated [18] that only 13% of the energy deposited by ionizing radiation into NaI(Tl) is emitted as luminescence and fluorescence, with the remaining ~86% available "to fuel complex avenues of long-term energy storage [18]." Like the effects of UV exposure described in the Saint-Gobain technical note [25], this stored energy can lead to events looking akin to irradiation with ~keV electrons. Low-energy signals in the DAMA-LIBRA experiment require only 0.15% of the available energy deposited by muons[18]. Energy-conversion efficiency for materials used in TSL dosimeter reaches upwards of 13% in many cases [36], so David Nygren suggestion - that yearly modulation of low-energy background seen in NaI(Tl) can be caused by modulation of muon flux and possibly other environmental effects- cannot be rejected without more careful investigation. One needs to keep in mind that not only muon flux and residual radioactivity in NaI(Tl) crystals can affect COS-like dynamics, but the presence of trace impurities and crystal defects, and history of mechanical and thermal treatment of the samples (see fig. 4. B); i.e., crystals with nominally the same composition but prepared by different technologies can demonstrate quantitatively different SOC-type effects.

**Surface and bulk effects in dual-phase (liquid-gas) Xe and Ar detectors** (*problems on surfaces and in bulk*).

Given the operation principles and design of the dual-phase detector (Fig. 4), multiple mechanisms may lead to energy accumulation, charge accumulation, and delayed electron and photon emission in these detectors. As dual-phase designs are chosen in numerous ongoing and suggested projects attempting to expand dark matter searches into lower-mass/smaller-energy thresholds [3, 37], expand detector sizes [38], and search for Solar axions [39], clarifying and mitigating the origin of parasitic signals in these devices, which become more costly with increasing size, must be a central priority. Thus, here we discuss the potential offending mechanisms, with emphasis on those that can lead to avalanche-like relaxation dynamics.

The incomplete extraction of electrons from the liquid phase into gas and trapping of electrons and ions in an electrostatic potential trap at the liquid-gas interface with a subsequent release, as well as electron trapping and release by electronegative impurities in liquid are the often-listed problems. Note, that the potential trap for electrons at the liquid-gas boundary of liquid Xe is about 0.8 eV [40], so some yet unidentified mechanism needs to be present to liberate electrons. Each of these mechanisms can lead to parasitic electron emission events and position- and time-correlations with previous strong ionization events—e.g., electrons trapped by impurity will slowly drift toward the liquid surface, and if liberated by S2 light from later event, or liberated by chemical ionization, or by thermal activation (for low electron affinity trap), the electron will show up at the XY position correlated with the initial ionization location. Here diffusion is producing some spread or smearing; macroscopic liquid flow due to convection, and waves on the liquid surface can produce additional corrections. Beneficially, as electron drift time in the liquid in dual-phase detectors is short (few tens to 300-700 µs in large detectors as LUX or Xenon 1T), such events as photo-effect or ionizations caused by S2 light can be excluded by delay (time veto) after S2 events. However, electrons detected long after the maximum electron drift time delay present a problem for vetoing strategies.

Indeed, the LUX, Xenon1T, and RED group (Russian Emission Detector) experiments all observed delayed single-electron events at the positions of past strong ionization events (S2 events) [40,41,42,43], and the observed delays were occasionally significantly larger than the maximum free-electron drift time. Such significant delays are strong evidence for long relaxation or energy-storage processes. Observed increased electron emission lasts only for about a second after large ionization events in Xe detectors underground [42,43] and can be traced for up to 100 ms in small above-ground detectors [41]. The large rate of events prevents longer tracing in the above-ground experiments. Nonetheless, energy storage can be present for longer time intervals than 1 S.

Important sources of small ionization production in active volume of dual phase detectors are radioactive impurities decaying inside walls and electrodes or energetic particles interacting with walls and electrodes and producing a continuum of the event with "incomplete energy deposition" or "incomplete extraction" of produced ions into liquid. Fortunately, this background should be stable in time (if it is not connected to new radon coming into the detector) and can be modeled in a wide range of event energies – from few electrons to complete energy deposition in liquid with both S1 and S2 signals present [44]. Several other possible mechanisms can lead to the emission of few (1-8) electrons simultaneously, which are not sufficiently studied and difficult to model.

Metal and dielectric surfaces are a possible source of long-delayed single- and multiple-electron emission events. Field emission of electrons from metal becomes nonlinear and complex as one approaches breakdown conditions, but experimenters avoid operating detectors in these regimes. However, even when the electric field at the surface is far away from breakdown or absent and even when the probability of thermally excited electron emission is low because of low temperatures, multiple effects are known to produce electron and photon emissions. One family of these effects, the previously described TSEE, occurs when ionizing radiation pumps energy into materials. Other sources of energy as mechanical stress and dislocation motion in materials can trigger electron and photon emission from surfaces of dielectric and oxide layers on a metal surface; [45] describes such delayed electron emission from deformed metal. These effects allow researchers to visualize mechanical stress, fatigue, dislocation, and defect motion, etc., via electron microscopy and other scanning techniques (see, for example, [46]). The common feature across these examples is that multiple energy-transfer mechanisms can produce excited states on the surfaces and cause photons and electrons emission (see[24] for more discussion).

There is another important possibility: dislocation motion or grain re-arrangements can cause microscopic local changes of the structure of the oxide layer on the metal surface and lead to self-organization and subsequent distraction of microscopic electron-emitting structures on the surface capable to efficiently emit electrons in a low electric field.

Mentioned mechanisms are not yet sufficiently understood; using coatings with low secondary electron emission, like those in development for space microwave antennas [47], may provide an avenue to suppress these effects. Radiopurity of coatings containing niobium and titanium nitrides [47] can be a separate problem, but experimenting with coatings can help to separate and study another family of effects: energy accumulation in physisorbed solid layers on metal surfaces and interactions in between metal and physisorbed layers.

Up to 8-10 monolayers of solid physisorbed layers of Ar and Xe are present on all internal surfaces and electrodes in dual-phase detectors [48]. Our previous discussion on TSL and TSEE effects in solid Xe and Ar indicates that TSL and TSEE can be embedded into the design of dual-phase detectors—which consequently can explain how SOC-type dynamics in solid Xe and Ar can be present. In LUX, Xenon1T, RED-100, and many other detectors, the detectors' extraction grid is in liquid below ~5 mm of the liquid-gas interface, i.e., close to the gas-amplification region. A local gas electroluminescence event can consequently produce a localized imprint in the physisorbed solid xenon layer on the extraction grid due to photolysis [34]—wherein pairs of $Xe_2^+$ and $O^-$, $F^-$ ions are produced and trapped in the solid layer. Accumulation and avalanche-like recombination of trapped positive-negative ion pairs would produce delayed electron emission and correlated (multiple) electrons (and photons) emission events with characteristic timelines *independen* of the electron drift time in liquid, but still position-correlated with previous strong S2 emission event.

The Russian Emission Detector (RED) group performed a series of experiments to locate and suppress mechanisms of the delayed electron emission. For their small Xe detector operating above ground, they observed that the S2 pulse from strong ionization events (like muons) often was followed by another light production event at the perimeter of the detector, which they named S3 light pulse [43]. In their detector, the anode grid was in gas, but a metal rim of the grid holder was touching the liquid surface to collect non-extracted electrons. The interpretation was that a cloud of non-extracted electrons can travel over the liquid surface toward the detector perimeter where due to electric field shape electrons somehow can escape into the gas phase and produce the S3 light pulse by electroluminescence. Moreover, for single vertex ionization events they were able to trace in space and time a weak electron emission from the surface-bound electron cloud as it propagates from S2 position to S3 position on the practically straight line. In the LUX underground experiment similar multi-electron escape events arising at different x-y positions, compared to the anterior event, were observed after large ionization events [43]. They were named e-bursts. In contrast to the RED group observations, e-bursts mostly were observed close to the location of the preceding S2 event, but with longer delay after the S2 pulse – a few ms against a few hundred $\mu$s for S3 pulses in Russian experiments. Thus, unextracted electrons after large ionization events in the detector can escape to the gas from the liquid surface, but the exact mechanism is yet unknown. Without further experiments, one can speculate that differences in phenomenology observed in different detectors could be due to the presence of waves and impurities at the liquid surface.

In a 200 kg (about 160 kg active volume) Xe dual-phase detector operating above ground, the RED team installed an additional gate grid just under the extraction grid, which enabled the applicating a positive potential to this gate relative to the extraction grid to prevent electrons in the drift space (active volume) from reaching the liquid surface [49]. This closing potential was applied automatically for the duration of 1 ms whenever a luminescence signal S1 above a chosen threshold was detected. By applying this potential, RED prevented more than 95% of primary ionization electrons occurring during large events from reaching the liquid surface, which thereby reduced both the number of electrons that could be trapped on the liquid surface and the number of photons in the S2 pulse that could deposit energy on the grids and in the liquid volume. Application of this technique reduced the rate of single-electron events to about 250 kHz—compared with about 750 kHz in cases where counts during this 1 ms time interval were merely excluded. By observing that after deactivation of the stopping potential the single-electron signals are appearing with a delay corresponding to electron drift time from the second gate to the surface with no excess at the switching that could be attributed to the accumulation of electrons on the second gate, the RED group concluded that most of the remaining electron emission events originated below the extraction grid and the second gate, in the bulk liquid active volume and on the cathode. Thus, the RED-100 data suggest that one of the sources of multiple-electron emission events could be the bulk liquid, as most of the remaining (after double-gate activation) single-electron emissions originated in their experiments. [1]

To search for CEvNS events, the RED-100 detector will be installed under the reactor of Kalininskaya NPP, where the muon rate (which is believed to be the main ionization source in their set-up) will be ~5 times lower and, correspondingly, the single-electron rate is expected to be about five times smaller [49]. Even under the reactor, the high rate of single-electron events will lead to many random coincidences—so the CEvNS signals from reactor antineutrinos are expected to rise above random coincidences for about ~4 electron events [49]. The possibility of correlated emission of electrons, i.e., the presence of multiple electron emission events in an excess of coincidences of a single electron event was not yet studied by RED-100 and remains a seriouds unsertanty for the planned experiments.

---

[1] Potentially, arriving of positive ions to the cathode can cause electron emission, but there were no observations of long-delayed (more than 1 second) or re-appearing electron emission after large ionization events.

The presence of impurities in the bulk liquid likely plays a role here, and we can point to two distinct channels for producing single and multiple delayed electron-emission events in bulk liquids when impurities are present. The most obvious channel is due to the formation of slow atomic and molecular ions, and chemical radicals in presence of impurities. Slow ions, both positive and negative, move 1,000 to 10,000 times slower than free electrons [50] and will be present in liquid for a few seconds after ionization events. Chemical radicals (broken chemical bonds; fragments of larger molecules) can be formed by primary particles, by UV light, and by free electrons drifting in the strong electric field. Interactions between ions, radicals, and impurity molecules can produce delayed single-electron emission and few-electron emission events if the produced single electron liberates another one by impact during drift in the electric field. The oft-discussed possibility of electron capture and consequent release by electronegative impurity also falls under this group of processes. Because the formation of slow ions and radicals is correlated in space and time with primary events, delayed electron emission produced this way will also be correlated with primary ionization events.

The other channel relates to the possibility that impurities and radicals from molecular aggregates in liquid Xe and Ar, and these aggregates can accumulate sufficient energy to produce several electrons and photons simultaneously. This channel likely necessitates a brief discussion of experimental evidence for the "donor impurities" responsible for a variety of positive ions in Xe and Ar detectors. The liquid in dual-phase detectors continuously circulates by evaporating, passing as a gas through heated chemical getters, and condensing back into liquid. Thus, impurity concentrations stay at few ppb levels in liquid Xe and even lower in Ar detectors. While concentrations of electronegative impurities are usually monitored by measuring free-electron pulses decaying with travel distances in liquids, revealing the presence of positively charged impurity ions requires different techniques. In experiments [50,51], the mobility of $O_2^-$ ions and Xe ions ($Xe^+$ and $Xe_2^+$) was measured in a strong electric field. Normally, ionizing a liquid in a slab between two flat electrodes using high-energy X-rays should lead to the homogeneous production of ions and electrons within the liquid. Accordingly, the drift of each ion type should result in the current's pulse linearly going to zero as all the respective ions reach the electrode—so the total response should be composed of linearly decreasing current fragments. In the presence of impurities, positive Xe ions will collect electrons from impurities with smaller ionization potential (donors), which will result in visibly nonlinear current fragments in the relaxation response. The experiments revealed the presence of donor impurities. Importantly, in experiments [50, 51], the Xe gas was passed through a heated getter of the same type as used in all modern dual-phase detectors before condensing in the drift cell, but this procedure appears to have been insufficient to remove the donor impurities to a desirable level. Additional ionization cleaning—prolonged irradiation of the drift cell with X-rays (there were no plastics in the cell design)—resulted in the removal of impurities and the restoration of the expected linear patterns among the fragments in the response current. By introducing a controllable concentration of organic impurities, the authors of [50, 51] checked whether the free drift of Xe ions over 5 mm distance without noticeable charge exchange with impurities requires donor impurity concentrations below 0.1 ppb. Authors of [50, 51] supposed that donor impurities were due to the remains of hydrocarbon and fluorocarbon lubricants used in the gas-processing equipment, a factor that would still be in play for the dual-phase detectors.

Charge exchange (electron and proton exchange) in between ions (like Xe ions getting electrons from the donor impurities) is taking place in gases and have practical applications: it allows detection of trace impurities in Ar and Xe at ppt (part per trillion) level in atmospheric pressure ionization mass-spectrometers (APIMS) [52]. Delayed electron and photon emission due to the presence of impurities in noble gases also were studied and are known under the name afterglow [53]; studies of electrical breakdown in gases also reveal effects of long-living excitations and radicals both in gas and on surfaces of electrodes [54].

Positive ions arriving on the cathode and extraction grid in liquid Xe and Ar detectors are not Xe or Ar ions, but rather donor impurity ions. The kinetic energy of atomic or molecular ions drifting in an electric field in liquid is low, so only chemical energy (recombination energy) of these ions can cause electron emission (secondary electron emission) from electrodes. Dependence of secondary electron emission and multiple electron emission on chemical energy of slow ion recombination on the metal surface was studied in [55]. Results of this study may explain why ion recombination on cathode or extraction grid is not producing clear observable electron emission effects: chemical energies of "donor impurities" ions are too low to cause secondary electron emission. It is not possible to predict if the accumulation of the donor impurity ions or neutral donor atoms or molecules in solid physisorbed layers at the cathode surface can change electron emission from the cathode, but Malter effect [56] provides an example that electron emission can be strongly affected by the presence of positive ions in/ on thin dielectric film on metal surface.

We posit the existence of another channel producing delayed multiple-electron and/or -photon events specific to liquid gases (and much less probable in the gas phase): the formation of molecular aggregates/clusters with excessive chemical energy. This hypothesis is based on known examples. Liquid noble gases—especially Xenon, which can be liquified at rather high temperatures—have long been considered a convenient solvent for chemical and photochemical reactions [57]. Liquid Xe can stabilize chemical radicals produced by UV light (they can leave longer than in the gas phase) and make it possible to replace atoms or functional groups in organic molecules [58]. Also, impurities in liquid gases can make molecular aggregates that do not exist in the gas phase [58], and this effect is not just due to low temperature but also is due to "encaging effects" [58]—like stabilization of reaction intermediates in a solid rare gas matrix [35]. Examples are given in [58], wherein molecules reacting via exothermal reactions in the gas phase can form a low-energy hydrogen bond to produce a (meta-)stable complex, with the oscillation frequencies around this bond remaining measurable. If several molecules or radicals form meta-stable clusters with sufficiently large chemical energy, chain reactions in these clusters can release several electrons and photons.

The most recent examples of energetic stable molecules containing Xe are observations of the H-Xe-O-H and H-Xe-O-Xe-H molecules formed in solid Xe under the action of UV light in presence of water impurity [59,60]. Authors of [60] discuss that the lightest molecule containing two Xe atoms also has very high chemical energy- 8.4 eV and that $(Xe-O)_n$ polymer network could be possible. Thus, questions are if these molecules and polymers can be formed in liquid Xe (or in solid physisorbed Xe layers) and if the decay of these energetic molecules can result in auto-ionization/ free-electron production.

Another possibility for energetic aggregates production is the formation of stable clusters by impurity molecules and the production of excitations in these clusters due to ionization events in the detector. It was recently discovered that cold clusters (droplets) formed by He atoms in the presence of several electronic excitations (i.e., excimer molecules) can result in spontaneous emissions of several free electrons, and the probability of this relaxation channel is higher than other channels [61]. The study's authors speculate that similar effects may be present in clusters of other atoms.

Water molecules are known to form hydrogen bonds, and water clusters can be present in liquid Xe or Ar. It is reasonable to expect that water clusters can be floating on the liquid Xe or Ar surface. These clusters can attach one or several unextracted electrons which also are bounded to the liquid surface. Charged claster can be easier extract to the gas, where it can evaporate and release electrons, UV light can produce excitations in these clusters or energetic molecules like in [59,60]. Decay of such complexes can result in the emission of several electrons.

Thus, one can think of multiple mechanisms which can lead to delayed electron and photon emission, including multiple electrons or photons emission, in dual-phase Xe and Ar detectors. These emission events can originate in the bulk liquid, on the surfaces of the electrodes—either in solid physisorbed layers on surfaces, or in a metal oxide layers—or on the liquid-gas boundary. Current understanding of involved condensed matter and chemical processes is currently insufficient to predict the relative importance of different mechanisms.

The possibility (at least hypothetical) of chain reactions, impurity cluster formations, and SOC-type dynamics in dual-phase Xe and Ar detectors and scintillation liquid Xe and Ar detectors leads us to several conclusions. First, delayed multiple photon-emission events in liquid Xe and Ar scintillation detectors could be possible—like observed in a NaI(Tl) scintillator after UV exposure. A spectrum of low-energy events peaking toward zero energy was observed in experiments where the effectiveness of PTFE UV light reflectors was evaluated [62] by detecting scintillation of liquid Xe excited by a $^{241}$Am source (see Fig. 6). Multi-photon events in this "zero-energy" peak may be due to relaxations in energetic clusters in the bulk liquid and due to formation under UV irradiation and subsequent decay of exciplex complexes (pairs of $Xe_2^+$ and $F^-$ ions) or "water complexes" (energetic molecules like H-Xe-O-H ) in solid physisorbed films on surfaces, and solid-like Xe in pores in PTFE. Interestingly, at low temperatures UV irradiation can cause delayed luminescence from the solid physisorbed Xe films on surfaces and in PTFE pores when only Xe gas is present – as condensation of Xe into pores and formation of solid layers on surfaces is not required the presence of liquid.

Second, in dual-phase detectors, delayed events with simultaneous emissions of a few photons and electrons could be possible. For these events, both S1 and S2 lights could be observed. Formations of clusters with large chemical energies would be rare "chemical events," but the number of "excessive" low-energy electron-like recoils observed by Xenon1T over its runtime is also small [39]. In other words, detection of rare chemical events may be as difficult as detection of dark matter particles, but any dark matter nuclear recoil candidate event should at least be suspected to be a rare chemical event.

Importantly, in low energy neutron scattering experiments, when detection of neutrons scattered in the Xe (Ar) dual-phase detector active volume is possible with backing detectors, one can select a class of events where no energy depositions were detected during 30-100 ms before a neutron scattering event, so the efficiency of electron production by low energy recoils of Xe

(Ar) atoms in liquid can be calibrated up to single-electron accuracy [63]. This calibration can be used for example to measure the low energy nuclear recoil spectrum for recoils caused by neutrinos from a spallation source when a strobe signal for the neutrino appearance is present [64]. Interestingly, pre-selection of neutron scattering events in the detector with no other events before them [63] removes events "spoiled" by the relaxation processes from prior ionization events and allows to see 1, 2, 3, 4 electron events as clearly separated peaks in the spectrum of events. When the timing of the event is not known, it is not yet clear how one can distinguish random low-energy interactions with particles from parasitic delayed electron emission events.

On the other hand, the authors are optimistic about lowering the parasitic electron emission rate. Most of the mechanisms for the delayed electron emission discussed above are extrinsic: impurities are required in bulk or on surfaces to cause delayed electron emission. Removing all plastics (including PTFE) from the detector active volume and making the detector bakeable to effectively remove water films before cooling detectors seems to be reasonable steps to try. Cleaning gases to PPT level of impurities using getters was demonstrated [52], but ionization cleaning of liquid (like in [50,51]) can also be used *in sity* in a separate cold volume to avoid damage of photon detectors. An analog of APIMS can be arranged in liquid by letting ions produced in liquid exchange electrons with impurities during drift path in liquid before ions with liquid enter a high vacuum part containing mass-spectrometer, where liquid droplets evaporate. This way one can also detect weakly-bounded ions which can decay if heated in gas to room temperature. Another technique to detect impurities *in situ* could be spectroscopy of "delayed luminescence" i.e., spectroscopy of light produced in liquid by the strong ionization event or controllable electric discharge after a sufficient delay to enable free electrons and excimer Xe or Ar molecules to dissipate.

While effects of wire material and wire conditioning on electron emission in dual-phase detectors were extensively studied (studies for LUX are the most recent example [66] ), there were no attempts to separate effects and processes in metal and metal oxide from effects and processes in solid physisorbed layers on the wire surface. Some observations – like periods of strong electron emission when voltage is first applied to the detector or re-applied after a sufficiently long break - are pointing to the existence of such effects. Wire coating with low secondary electron emission conducting material [47] was not yet tried. An initial check for effects related to structures formed in a physisorbed solid layer on wires due to ion and impurity accumulation can be done by the application of strong AC voltage in between cathode wires.

The Author believes that additional efforts in detector physics research and broader cooperation with the condensed matter physics community will help to avoid misinterpretations and resolve background problems to accelerate future discoveries.

**Superconducting detectors.**

There is a growing number of projects using superfluid, superconducting, and other low-temperature, solid-state detector technologies to continue the quest for dark matter particles and coherent scattering of low-energy neutrinos on nuclei. Though we cannot review this fast and dynamically growing field of research in this paper—interested readers may pursue [1] or read information about current projects in the materials of the Magnificent CEvNS workshop 2018 [67], 2019, and 2020 (published online)—here, we wish to point that searching for events where energy deposited by the external particle in the detector is in 1-10 eV range and even in meV range can also be limited by SOC-like dynamics in materials. The appearance of unexpected low energy background in 1-10 eV range in many types of low energy threshold cryogenic particle detector was reported and discussed on EXSES workshop in summer of 2021 [68].

At low temperatures, where photon detectors with energy threshold in the meV range operate, many subsystems in materials demonstrate complex and history-dependent relaxation properties. As illustrations, we can mention charges localized on boundaries and interphases in SQUIDs [69], or magnetic moments of impurities in superconductors [70]. Glass-like relaxation processes mean that energy will be accumulated in materials due to any variations of electric or magnetic field—due to signals intentionally applied to the cold device or leaking to the experiment from the hot environment. And as different non-equilibrium configurations of charges, magnetic moments, nuclear moments, defects, etc. are interacting directly and indirectly (through lattice deformations—i.e., via phonons—or electron system), the authors posit avalanche-like relaxation events could also be possible. While these effects can be detrimental for low-temperature detectors and quantum information devices, this discussion goes far above the scope of this paper.

**Discussion.**

In a paper analyzing attempts to model scenarios of spontaneous symmetry breaking during early universe cooling by experiments with superfluid He3 overheated above superfluid transition temperature by energetic particles [71], Anthony Leggett remarked, "It will be all too obvious … that our current theoretical understanding of the processes which arise following the incidence of a high-energy particle on superfluid $^3$He is, to put it charitably, incomplete." Note that the author is talking here also about the recombination of ions and temperature equilibration processes in normal He3 fluid after inhomogeneous heating by particles. In keeping with Leggett's observation, searches for rare low-energy interactions with particles can easily become inconclusive because of insufficient understanding of condensed matter and chemical processes in detector materials.

We would like to stress here the connections of the background problem in low-energy threshold detectors with non-equilibrium thermodynamic ideas and concepts developed to describe systems under energy flow—such as those introduced by Ilya Prigogine: the appearance of dissipative structures, the appearance of order from chaos, and the generation of complexity [72]. These effects are present in live and non-live systems and are of fundamental importance for understanding the origin of life and the complexity of live systems. For detector development, people are interested in the systems with the simplest energy relaxation passways (noble liquids and gases are a good example), where unambiguous interpretation of experimental results is possible. It is not yet feasible to formulate the general criteria to predict the complexity of the dynamics in a driven non-linear system; this is also a problem for SOC or SOC-like dynamic. This means that in each case one needs to study the dynamic of the system in question. Generation of complexity can be rephrased as the possibility for different channels of energy accumulation to interact and this way to produce a more complex relaxation dynamic; this calls for experimental techniques to disentangle different mechanisms.

SOC-like scenarios and other dynamical effects in driven systems feasibly can explain lingering annoyances in dark matter searches and in attempts to detect coherent scattering of a low-energy reactor or solar neutrinos. When energy accumulated in materials releases via avalanche-like events, the resulting bursts of photon and electron emission problematically mimic interactions with external particles - this scenario demonstrates the common problem. One of the goals in writing this paper was to raise awareness in the particle physics community that the absence of low-energy gammas, electrons, or neutrons cannot guarantee the absence of low-energy background events with energies above thermal- when long-living hot excitations can be produced in the detector materials. The presence of such excitations makes possible the internal detector material mechanisms of single and multiple electrons or photons generation, even when the number of produced long-living excitations is low.

The dynamic properties of atomic or molecular clusters and other processes discussed in this paper are multi-particle quantum mechanics problems where first-principles calculations and modeling are difficult. This justifies using analogies with known effects and calls for experimental techniques to study energy accumulation and relaxation pathways. In this sense condensed noble gases and detectors are good model systems for studying such effects. The energy range of excitations of interest here can be accessed with absorption and emission spectroscopy and many other techniques; "exotic" or otherwise rare excitations in other systems, like cryogenic detectors at temperatures below 1K are much more difficult objects for experiments.

The fact that sources of parasitic low energy background are not yet unambiguously identified, though the problem was present already in the early dark matter particles detection experiments, like Xe10 [13], is pointing to the lagging of the detector physics and related condensed matter research from the requirements to present-day particles physics.

Finally, the author wishes to raise a potential bridge for this dark matter and neutrino detector-driven science into the domain space of superconducting devices and qubits. Given our expectation that SOC-type dynamics can take place in materials where interacting metastable states are present, we propose that non-thermal noise and decoherence effects may signal relaxation events within the metastable states and configurations that are present in materials at low temperatures. Our general expectations are that choosing materials with a lower number of low-energy states and operating detectors in a way that minimizes energy depositions in materials should universally lead to lower noise and a lower rate of parasitic events. While this short discussion necessitates further study far beyond the scope of this paper, we raise the point to illustrate that future cross-pollination may yield significant results for all these fields.

**Acknowledgments.**
Author needs to thank Adam Bernstein (LLNL) for introducing him to the problem of background in low energy threshold dark matter and coherent neutrino scatter detectors, interest, and numerous discussions in the course of this work; to thank D.Yu. Akimov and A. I. Bolozdynya (Moscow RED group) for multiple insightful discussions on the physics of dual-phase detectors (emission detectors in Russian literature); to thank prof. David Nygren (U. Texas) and prof. Frank Calaprice (Princeton) for discussions and comments on earlier versions of the manuscript; and special thanks to Michele Rubin for the help with editing this paper.
This work was performed under the auspices of the U.S. Department of Energy by Lawrence

Livermore National Laboratory under Contract DE-AC52-07NA27344. Authors acknowledge LDRD grant 17-FS-029 and 20-SI- 003, and DOE field work proposal number SCW1508. LLNL-JRNL-824835-DRAFT

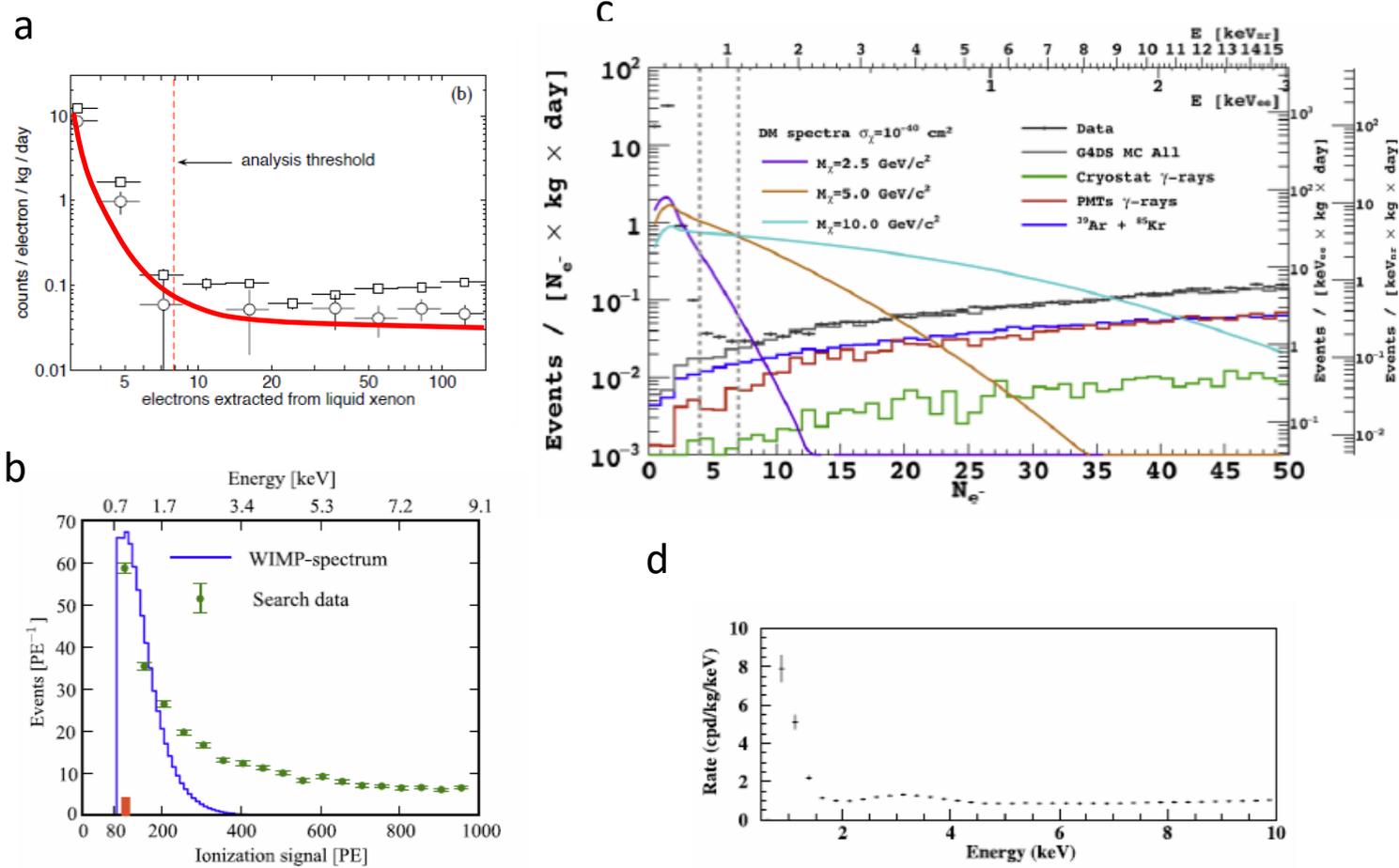

Fig. 1. Dark matter particle detectors operating underground (low background). A: Xenon 10 experiment [13], 10 kg liquid Xe TPC; analysis of electroluminescence signal only; we add red curve to illustrate contribution we expect due to energy accumulation effects in materials. B: Xenon 100 experiment [14], 100 kg liquid Xe TPC, analysis of electroluminescence signal only; the ~20 photoelectrons registered by PMTs in this experiment correspond to 1 electron extracted from the liquid; excessive few-electron noise is present, but "pre-breakdown electron multiplication" was observed in a too low electric field, so additional mechanisms for background can be present. C: Dark Side 50 experiment [6], 50 kg liquid Ar TPC; small signals (below ~5 electrons) are excluded from this analysis; authors admit that they do not understand the excessive background in the range of 4–7 electron events; they believe that the 1–4 electron events are coincidences of single electrons, but these events' energy resolution and trigger efficiency at 1 electron are insufficient to draw this conclusion. D: DAMA-LIBRA experiment, NAI(Tl) scintillator, energy deposition of 1 keV results here in registration 5.5-7.5 photons by PMTs; figure from paper [18] where David Nygren was discussing background data from Phase 1; pictures in Phase 2 analysis papers like [16] are shoving data where lowest energy portions of background were removed from the analysis and plots.

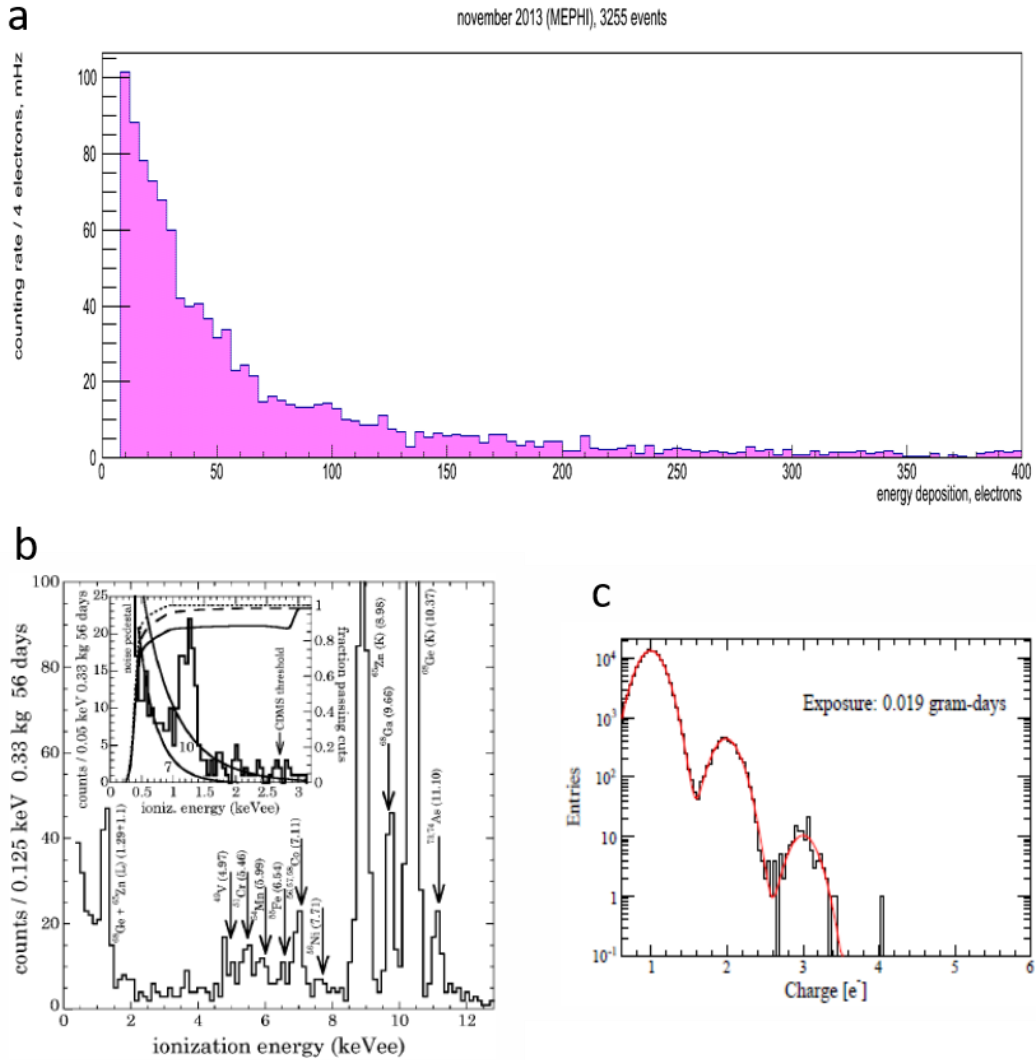

Fig. 2. Examples of background in surface experiments. A: RED1 Xe detector operating in old reactor building, courtesy of Dmitri Akimov; (the detector design,is described in [41]); B: High purity Germanium detector [8] operating near power reactor (search for coherent neutrino scattering signal). C: SENSEI project [17], thick silicon CCD, spectrum of small events, frames containing strong ionization events and frames after strong ionization event are rejected.

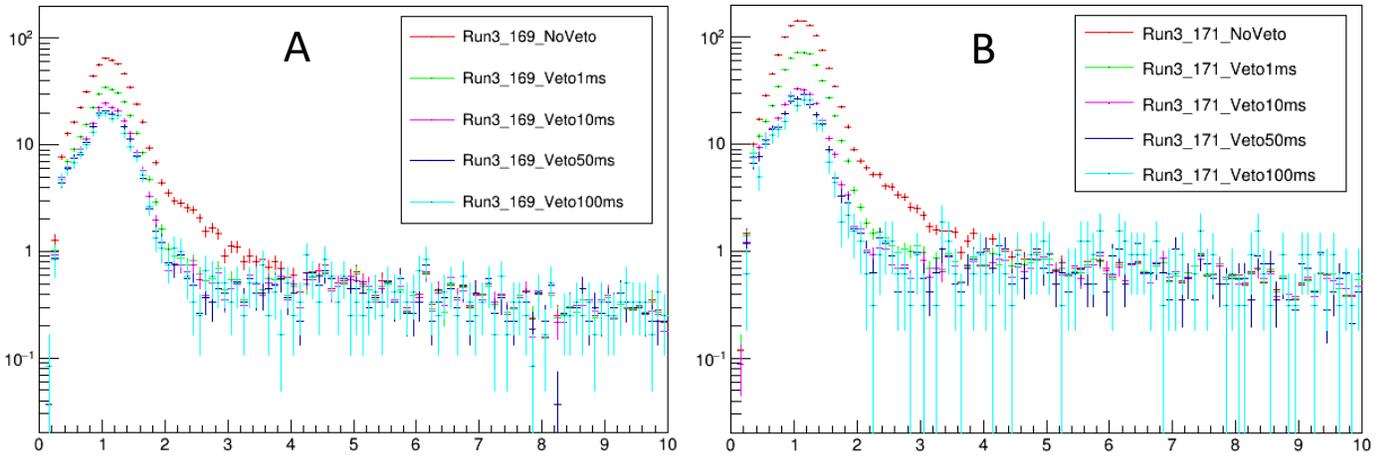

Fig. 3. Effect on low energy background from Co60 source placed outside detector and collimated with led blocks to expose limited region inside active liquid volume. In calculating spectra events immediately after large events (more than 500 e) during veto period was ignored and corresponding correction was done to active time. A.- Rate of high energy events (more than 500 e) is 13 Hz. B- Collimated Co60 source is present, rate of high energy events (more than 500 e) is 21 Hz.

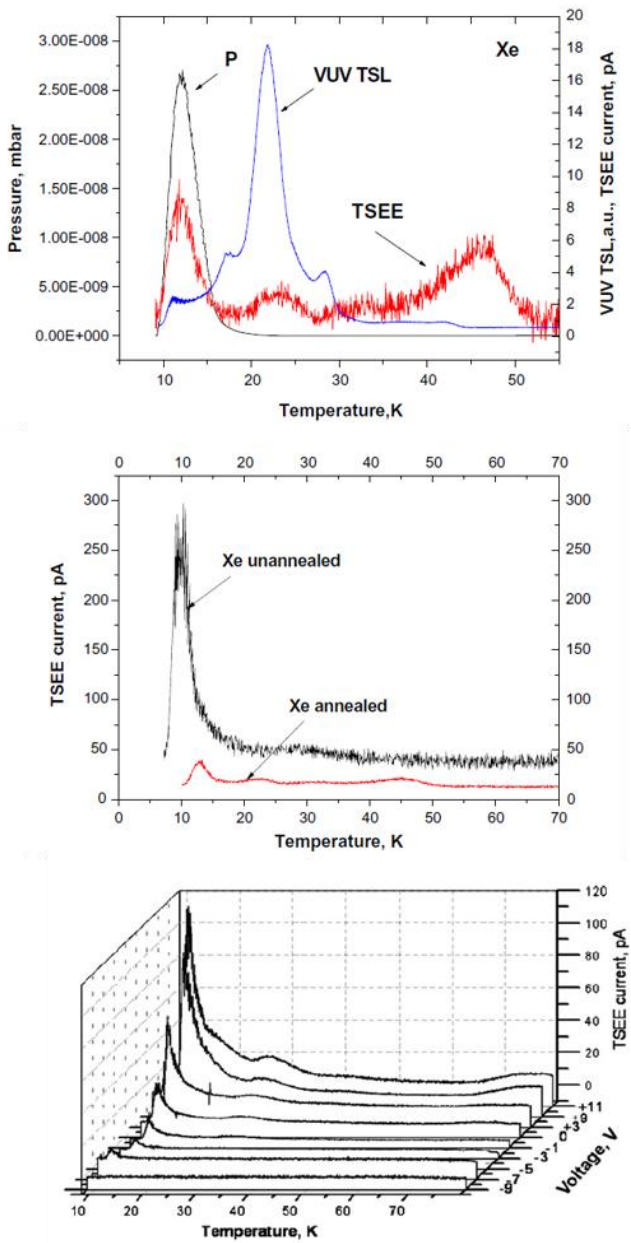

Fig. 4. Data from the paper [27,28] of Savchenko et.al. A: Thermally stimulated luminescence, thermally stimulated electron emission and thermally stimulated sublimation (in units of pressure increase) for solid Xe crystal irradiated at 6 K temperature with electrons. B: Suppression of thermally stimulated electron emission for Xe crystal annealed at higher temperature before irradiation at 6 K. C: Thermally stimulated electron emission current measured with different potential bias of electrometer current collecting electrode with respect to experimental chamber, retarding potential ~ 8 V is required to suppress current.

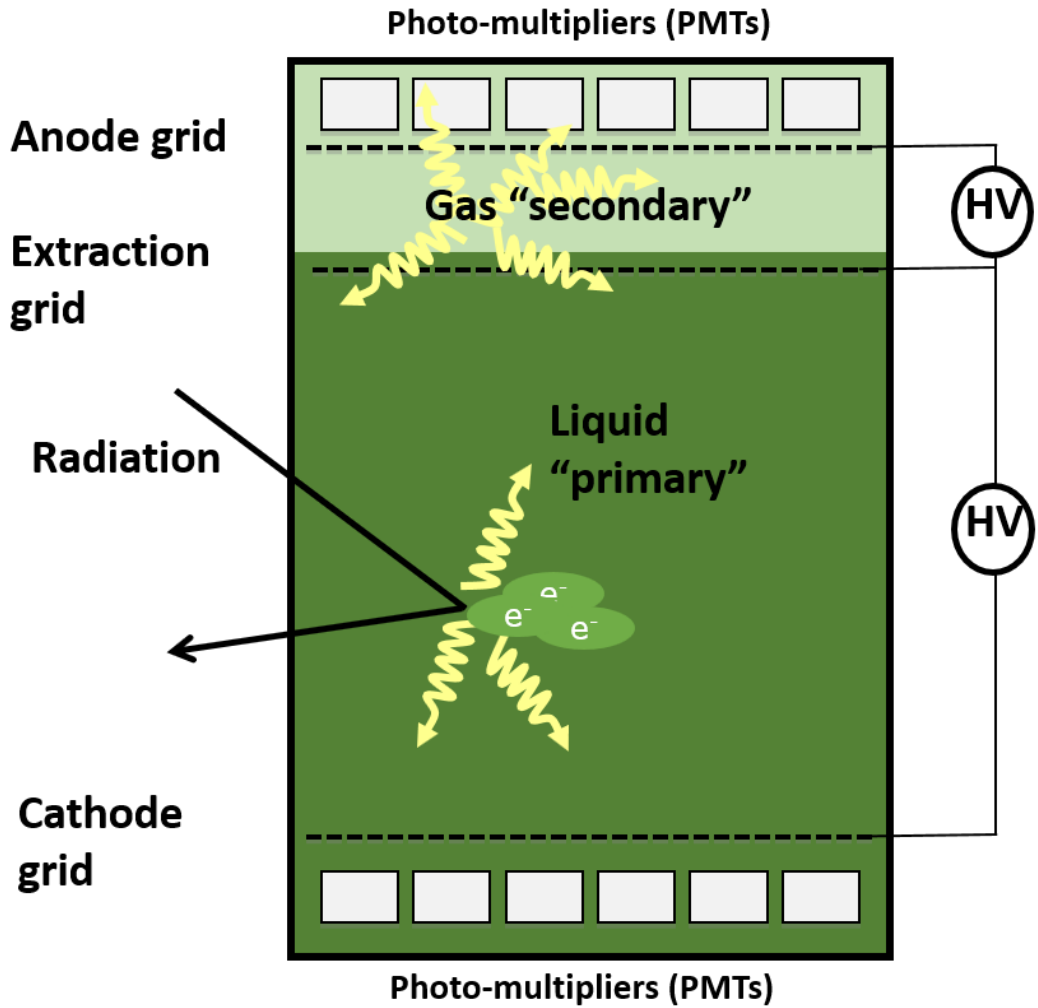

Fig. 5. Operation principles of the Dual-Phase Detector (noble liquid TPC). Electron-ion pairs produced in liquid active volume due to an interaction with primary particle are separated by drift electric field; primary luminescence (S1) signal is detected by upper and lower arrays of PMTs. Electrons are driven up toward liquid-gas interface, where they are emitted into gas and produce proportional gas electroluminescence (S2 signal) on their way to anode. Center of mass of S2 signal determined with the upper PMT array (above anode grid) provide invent X-Y position information; total S2 signal also provide the total number of electrons extracted from the liquid into gas. Delay in between S1 and S2 signals gives electron drift time in liquid and provide Z coordinate of the primary event (dual-phase detector is sub- type of time -projection chambers).  For low energy events primary scintillation becomes undetectable because of incomplete surface coverage with photodetectors and not very high photon detection efficiency.

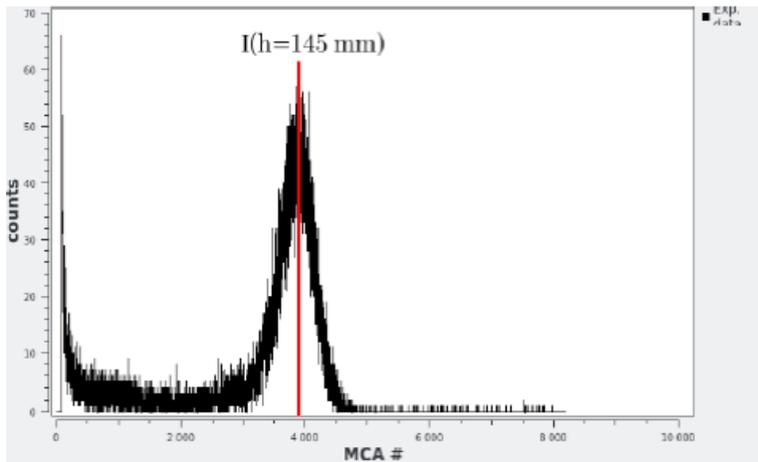

Fig. 6. Example of experimental spectrum obtained using PTFE reflector sample in liquid Xe and $^{241}$Am source that emits alpha particles with an energy 5.486 MeV [62]. Particles has ~50 mm range in liquid Xe and produce ~3.36 x $10^5$ UV photons in liquid. These photons interact with impurities in bulk liquid and on surfaces and as PTFE is porous, can cause PTFE - Xe infractions in near-surface PTFE layer. The few photons peak is result of delayed photon emission- i.e. result of all energy relaxation processes in the system